\documentclass[aps,preprint,prl,superscriptaddress]{revtex4}

\usepackage{graphicx}
\usepackage{amsmath}
\usepackage{amssymb}
\usepackage{units}
\usepackage{color}

\usepackage{setspace}

\begin{document}

\title{Phonon-pump XUV-photoemission-probe in graphene: evidence for non-adiabatic heating of Dirac carriers by lattice deformation}

\author{Isabella Gierz}
\email{Isabella.Gierz@mpsd.mpg.de}
\affiliation{Max Planck Institute for the Structure and Dynamics of Matter, Hamburg, Germany}
\author{Matteo Mitrano}
\author{Hubertus Bromberger}
\affiliation{Max Planck Institute for the Structure and Dynamics of Matter, Hamburg, Germany}
\author{Cephise Cacho}
\author{Richard Chapman}
\author{Emma Springate}
\affiliation{Central Laser Facility, STFC Rutherford Appleton Laboratory, Harwell, United Kingdom}
\author{Stefan Link}
\author{Ulrich Starke}
\affiliation{Max Planck Institute for Solid State Research, Stuttgart, Germany}
\author{Burkhard Sachs}
\affiliation{I. Institut f\"ur Theoretische Physik, Universit\"at Hamburg, Germany}
\author{Martin Eckstein}
\affiliation{Max Planck Institute for the Structure and Dynamics of Matter, Hamburg, Germany}
\author{Tim O. Wehling}
\affiliation{Institut f\"ur Theoretische Physik, Universit\"at Bremen, Bremen, Germany}
\author{Mikhail I. Katsnelson}
\affiliation{Institute for Molecules and Materials, Radboud University Nijmegen, Nijmegen, The Netherlands}
\author{Alexander Lichtenstein}
\affiliation{I. Institut f\"ur Theoretische Physik, Universit\"at Hamburg, Germany}
\author{Andrea Cavalleri}
\affiliation{Max Planck Institute for the Structure and Dynamics of Matter, Hamburg, Germany}
\affiliation{Department of Physics, Clarendon Laboratory, University of Oxford, Oxford, United Kingdom}

\date{\today}

\begin{abstract}
We modulate the atomic structure of bilayer graphene by driving its lattice at resonance with the in-plane E$_{\text{1u}}$ lattice vibration at 6.3\,$\mu$m. Using time- and angle-resolved photoemission spectroscopy (tr-ARPES) with extreme ultra-violet (XUV) pulses, we measure the response of the  Dirac electrons near the K-point. We observe that lattice modulation causes anomalous carrier dynamics, with the Dirac electrons reaching lower peak temperatures and relaxing at faster rate compared to when the excitation is applied away from the phonon resonance or in monolayer samples. Frozen phonon calculations predict dramatic band structure changes when the E$_{\text{1u}}$ vibration is driven, which we use to explain the anomalous dynamics observed in the experiment.
\end{abstract}

\maketitle

Optical excitation of Dirac carriers in graphene has to date been known to occur through two mechanisms. For photon energies higher than twice the chemical potential ($\hbar\omega_{\text{pump}}>2|\mu_e|$), direct interband excitation takes place \cite{Johannsen_2013,Ulstrup_2014}, resulting in population inversion for sufficiently high fluences \cite{Li_2012,Winzer_2013,Gierz_2013}. For doped samples and lower photon energies ($\hbar\omega_{\text{pump}}<2|\mu_e|$), the Dirac carriers are heated by metallic free carrier absorption \cite{Winnerl_2011,Gierz_2013}, where the peak electronic temperature is determined by the pump fluence \cite{Gierz_2013}.

\begin{figure}
	\center
  \includegraphics[width = 0.5\columnwidth]{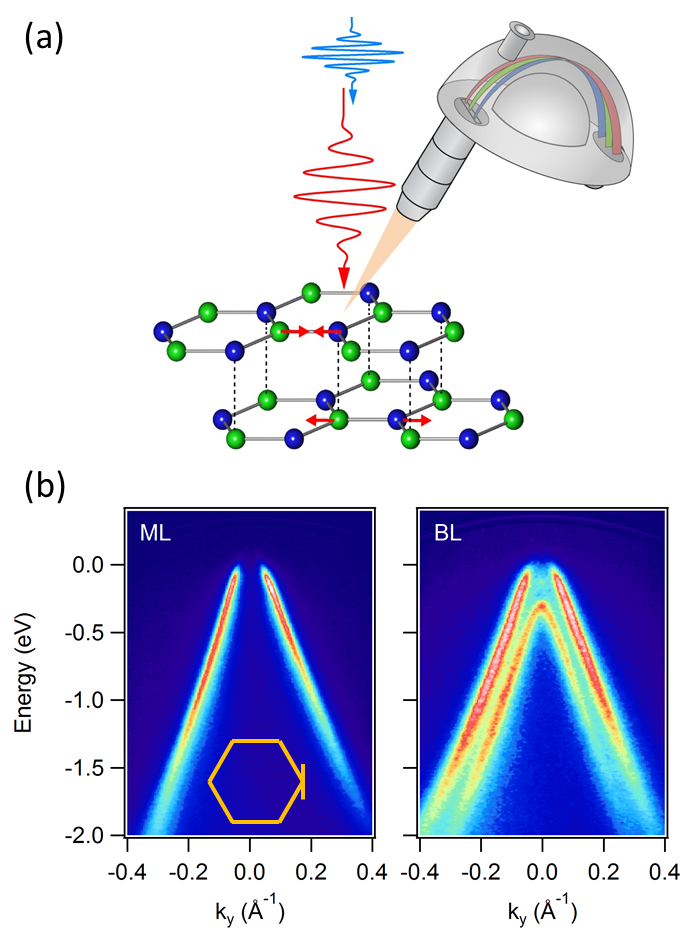}
  \caption{(a) Sketch of the tr-ARPES experiment. The pump pulse (red) at normal incidence resonantly excites the in-plane E$_{\text{1u}}$ lattice vibration in bilayer graphene. A collinear extreme ultra-violet probe pulse (blue) ejects photoelectrons that pass through a hemispherical analyzer and impinge on a two-dimensional detector. (b) Equilibrium band structure for hydrogen-intercalated monolayer (ML) and bilayer (BL) graphene measured with HeII$\alpha$ radiation for a cut through the K-point perpendicular to the $\Gamma$K-direction (see inset).}
  \label{fig1}
\end{figure}

Here, we introduce a new mechanism, active when an infrared optical field is made resonant with a vibrational mode that modulates the band structure. We show that anomalous heating of the Dirac carrier distribution occurs when the E$_{\text{1u}}$ phonon mode of bilayer graphene \cite{Ando_2007,Bonini_2007,Park_2008,Malard_2008,Ando_2009,Gava_2009,Ferrari_2013} is driven to large amplitudes with a coherent mid-infrared field. This mode is particularly interesting as it exhibits a pronounced Fano profile and large oscillator strength in the conductivity spectrum, originating from an anomalously strong coupling to electronic interband transitions \cite{Zhang_2009,Kuzmenko_2009,Tang_2010}. Further, the E$_{\text{1u}}$ motion of the two triangular lattice units (Fig. \ref{fig1}a) is expected to periodically open and close a band gap at the K-point \cite{Cappelluti_2012}, a prospect of general interest for the physics of graphene. Finally, the frequency of the mode is fast compared to the electron-phonon scattering time, resulting in a breakdown of the adiabatic Born-Oppenheimer approximation as proposed in the context of Raman scattering experiments for monolayer graphene \cite{Pisana_2007}.

Quasi-freestanding epitaxial graphene mono- and bilayers, grown on the silicon-terminated face of silicon carbide (SiC), are used in this work \cite{Riedl_2009} (for further details see \cite{SupMat}). The equilibrium ARPES spectra for these samples, measured at room temperature with HeII$\alpha$ radiation at $\hbar\omega=40.8$\,eV, are shown in Fig. \ref{fig1}b. These measurements were performed along a momentum cut that crosses the K-point perpendicular to the $\Gamma$K-direction (inset Fig. \ref{fig1}b), and show the characteristic $\pi$-band dispersion of graphene. Both samples are lightly hole-doped due to charge transfer from the substrate with hole concentrations of $n=6\times10^{12}$\,cm$^{-2}$ and $n=4\times10^{12}$\,cm$^{-2}$ in monolayer and bilayer graphene, respectively.

Mid-infrared pulses were used to excite these samples, either resonantly with the E$_{\text{1u}}$ mode in bilayer graphene ($\lambda_{\text{pump}}=6.3$\,$\mu$m) or at other wavelengths between 4\,$\mu$m and 9\,$\mu$m. The resulting non-equilibrium Dirac carrier distributions were tracked with tr-ARPES as a function of pump-probe time delay using extreme ultra-violet (XUV) femtosecond pulses at 31\,eV photon energy. The experiments were performed at 30\,K base temperature.

\begin{figure*}
	\center
  \includegraphics[width = 1\columnwidth]{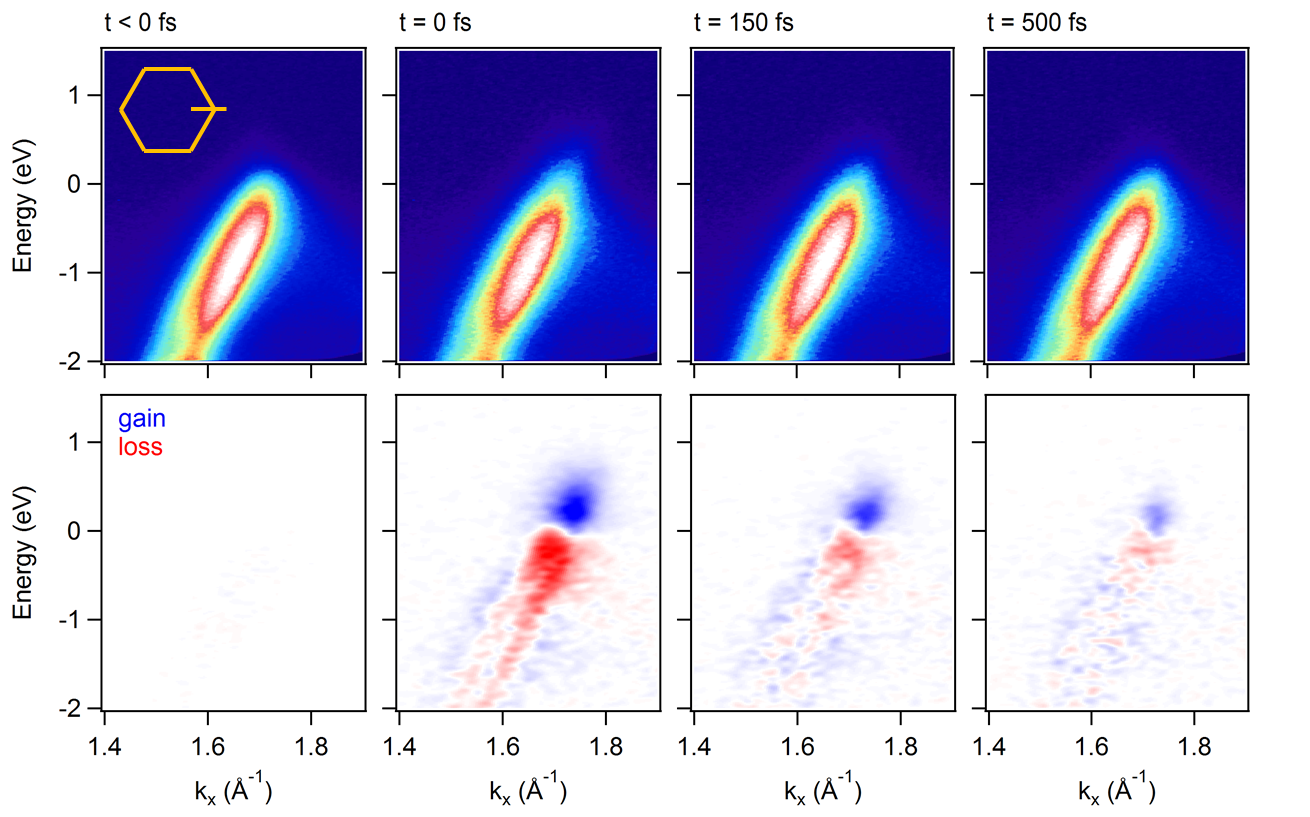}
  \caption{Snapshots of the electronic structure along the $\Gamma$K-direction (see inset) of bilayer graphene for different pump-probe delays (upper panel) together with the corresponding pump-probe signal (lower panel). The excitation wavelength was 6.3\,$\mu$m in resonance with the in-plane E$_{\text{1u}}$ lattice vibration. The fluence was F = 0.26\,mJ/cm$^2$.}
  \label{fig2}
\end{figure*}

In Fig. \ref{fig2} we present snapshots of the carrier distributions (upper panel) of bilayer graphene, along with the pump-induced changes of the photocurrent (lower panel) for different pump-probe delays after excitation of the E$_{\text{1u}}$ lattice vibration at $\lambda_{\text{pump}}=6.3$\,$\mu$m. These measurements were taken along the $\Gamma$K-direction, where only two of the four $\pi$-bands are visible due to photoemission matrix element effects \cite{Shirley_1995}. Zero time delay was set to the peak of the pump-probe signal.

\begin{figure}
	\center
  \includegraphics[width = 0.7\columnwidth]{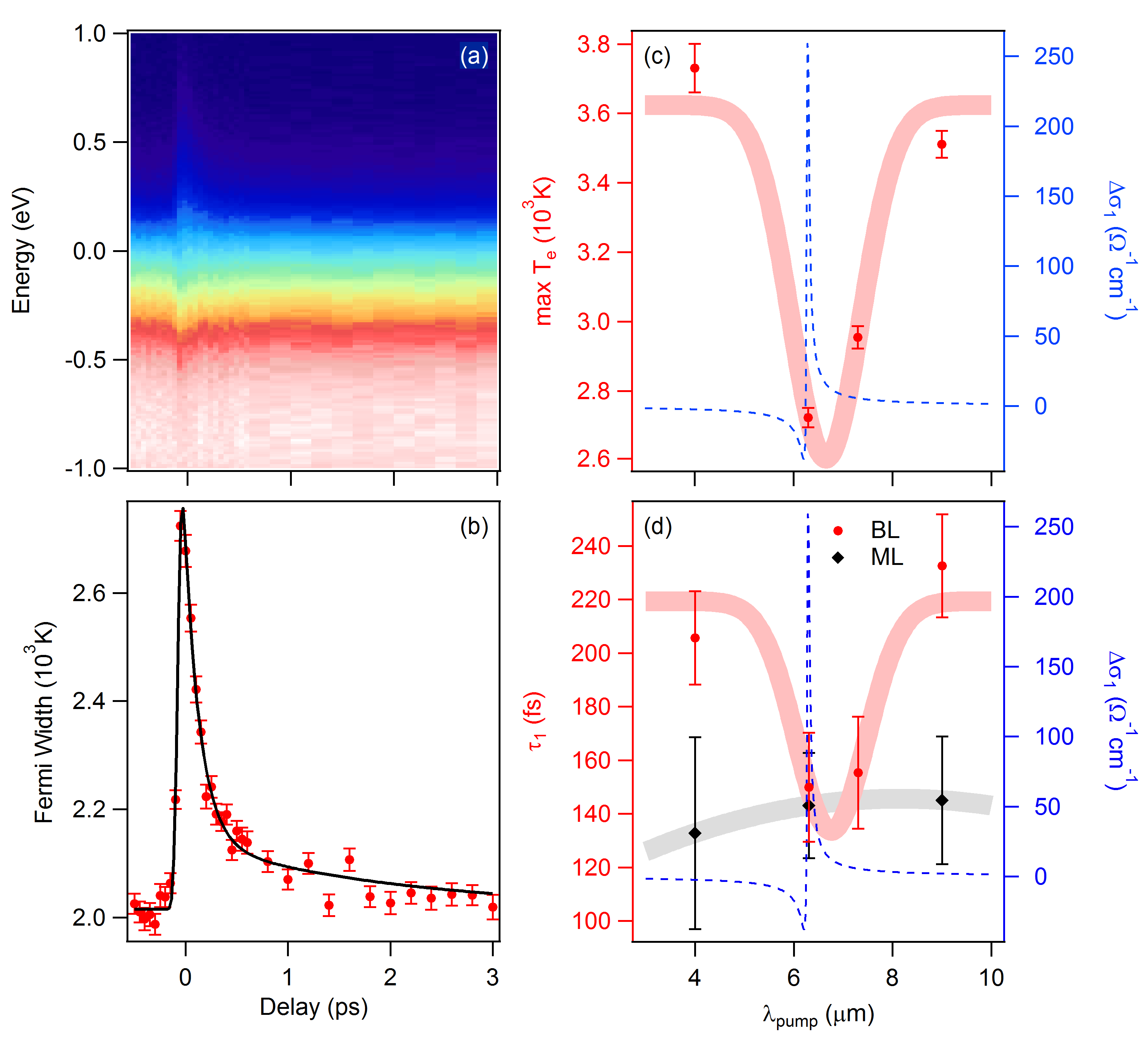}
  \caption{(a) Momentum-integrated photocurrent from Fig. \ref{fig2} as a function of energy and pump-probe delay in the vicinity of the Fermi level. (b) Width of the Fermi-Dirac distribution as a function of pump-probe delay (data points). The continuous line is a fit including an error function for the rising edge and the sum of two exponentials to describe the decay. (c) Dependence of the peak electronic temperature on excitation wavelength for constant excitation fluence of 0.26\,mJ/cm$^2$. (d) Dependence of the fast relaxation time $\tau_1$ on excitation wavelength for constant excitation fluence of 0.26\,mJ/cm$^2$. The corresponding data for monolayer graphene for fluences between 0.26\,mJ/cm$^2$ and 0.8\,mJ/cm$^2$ is shown in black for direct comparison. Continuous lines in (c) and (d) are guides to the eye. Dashed lines in (c) and (d) show the Fano line shape of the phonon in the real part of the optical conductivity $\Delta\sigma_1$ from \cite{Kuzmenko_2009}.}
  \label{fig3}
\end{figure}

Consistent with previous measurements performed in the free carrier absorption regime \cite{Gierz_2013}, we observe a broadening of the Fermi edge due to an increase in electronic temperature, without signature of population inversion.  The transient distributions of Fig. \ref{fig2} were integrated over momentum, and plotted as a function of pump-probe delay (Fig. \ref{fig3}a). At each time delay, the momentum-integrated photocurrent was fitted by a Fermi-Dirac distribution, to obtain the width of the Fermi cut-off, related to the electronic temperature, T$_e$. The time evolution of the Fermi width (Fig. \ref{fig3}b) is found to follow a double exponential decay, generally ascribed to the emission of optical ($\tau_1$) and acoustic phonons ($\tau_2$) \cite{Moos_2001,Kampfrath_2005,Dawlaty_2008,George_2008,Breusing_2009,Sun_2010,Kang_2010,Huang_2010,Lui_2010,Hale_2011,Breusing_2011,Gao_2011,Malic_2011}.

Experiments for different excitation wavelengths were performed at a constant pump fluence of F = 0.26\,mJ/cm$^2$ (see \cite{SupMat}). The wavelength dependence of the peak electronic temperature, max T$_e$, and the fast relaxation time, $\tau_1$, are plotted in Fig. \ref{fig3}c and d, respectively. Both quantities exhibit a strong anomaly at the phonon resonance, with a reduced peak temperature and faster relaxation time ($\tau_1$). The second relaxation time ($\tau_2 = 2.8\pm0.6$\,ps) is instead found to be independent of pump wavelength.

Comparison with monolayer graphene (see \cite{SupMat}), where the in-plane lattice vibration is not infrared active and hence cannot be excited with light \cite{Ferrari_2013}, clearly indicates that the anomalies observed in the bilayer system are not present. This is most obvious in Fig. \ref{fig3}d, where we added the $\tau_1$ values for monolayer graphene (black data points) which are found to be independent of pump wavelength.

\begin{figure}
	\center
  \includegraphics[width = 0.7\columnwidth]{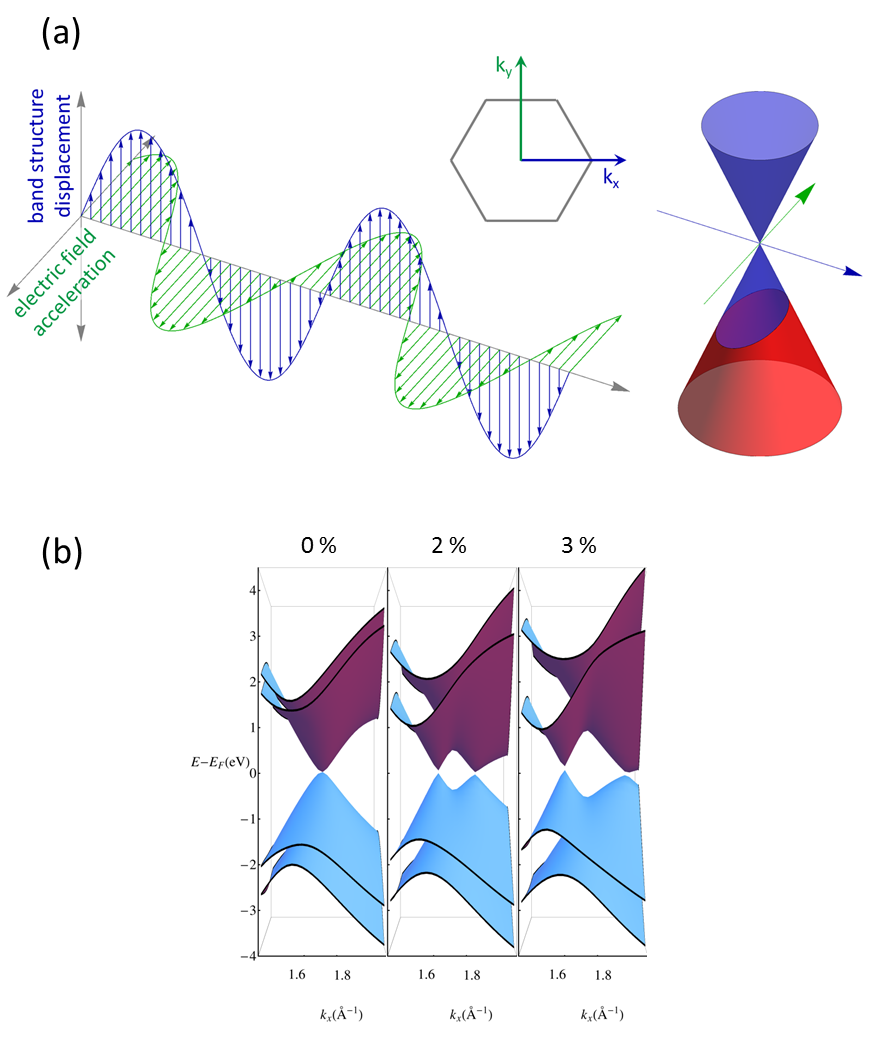}
  \caption{(a) Sketch of the excitation due to electric field acceleration (green) and band structure shift (blue). The field polarization lies along $k_y$ ($\Gamma$M). The displacement of the atoms in real space leads to a corresponding displacement of the Dirac cone in momentum space along $k_x$ ($\Gamma$K). The combined effect results in a complicated sloshing motion of the electrons that depends on the relative phase and amplitudes of the electric field acceleration and the band structure displacement, respectively. (b) Density functional theory calculations of the electronic structure around the K-point in the vicinity of the Fermi level for different lattice distortions of 0\%, 2\% and 3\% of the equilibrium lattice constant along the E$_{\text{1u}}$ mode coordinate.}
  \label{fig4}
\end{figure}

The data presented above points then to important differences between excitation on and off resonance with the in-plane lattice vibration, which is interpreted in the following.

In the case of free carrier absorption \cite{Gierz_2013}, the excitation is a result of periodic acceleration and deceleration of the carriers along the direction of light polarization, $k_y$ ($\Gamma$M-direction, green arrows in Fig. \ref{fig4}a). Carrier heating occurs in this case because the Dirac carriers scatter during this oscillatory motion.

When excitation is resonant with the E$_{\text{1u}}$ phonon, in addition to this time dependent polarization, one expects a modulation of the band structure, as the motion of the atoms makes the hopping between sites time dependent. Qualitatively, one can think of the motion of the atoms in real space directly translating into an oscillatory displacement of the Dirac cone itself \cite{Pisana_2007}. Note that this motion of the Dirac cone occurs along $k_x$ ($\Gamma$K-direction, blue arrows Fig. \ref{fig4}a), that is, perpendicular to the free carrier absorption occurring along $k_y$. Crucially, since the atomic motion is fast compared to the electron-phonon scattering time, the electrons do not follow the motion of the atoms promptly and experience a complex motion that breaks the adiabatic Born-Oppenheimer approximation \cite{Pisana_2007}.
 
To understand the physics at hand more quantitatively, we calculate the changes in the band structure for a static lattice distortion along the normal mode coordinate. For a pump field of approximately 1\,MV/cm, as employed in our experiments, we estimate atomic displacements of $\sim$5\,pm, $\sim$2\% of the equilibrium lattice constant (see \cite{SupMat}). In Fig. \ref{fig4}b, we report calculated band structures for different lattice distortions along the E$_{\text{1u}}$ normal mode coordinate. Dramatic changes are induced by the vibration, with a momentum-splitting of the $\pi$-bands at the K-point and a huge shift of the $\sigma$-bands at the $\Gamma$-point towards the Fermi-level (see \cite{SupMat}). The oscillatory motion of the Dirac cone known from the monolayer (Fig. \ref{fig4}a) affects the upper and lower layer of the bilayer in opposite directions. Together with the interlayer coupling this leads to a splitting of the $\pi$-bands along $\Gamma$KM that increases with increasing distortion (Fig. \ref{fig4}b). The non-equilibrium occupancy of the transient electronic states is determined by the complex non-adiabatic motion sketched in Fig. \ref{fig4}a.

To describe the experimental results we consider how the dynamical changes in the density of states (see Fig. 4 in \cite{SupMat}) are expected to affect the electronic temperature. In order to approximately estimate the change in peak electronic temperature due to the E$_{\text{1u}}$ lattice distortion, we assume that during the lattice distortion both the electron number N(DOS,$\mu_e$,T$_e$) and the entropy of the system S(DOS,$\mu_e$,T$_e$) are conserved (for further details see \cite{SupMat}). The chemical potential and the peak electronic temperature in the absence of a lattice distortion are known from off-resonance tr-ARPES data ($\mu_0\sim-200$\,meV, $T_0\sim3600$\,K). Based on the DOS in Fig. 4 in \cite{SupMat}, we obtain $\mu_1\sim-140$\,meV and $T_1\sim2745$\,K for an atomic displacement of 3\% of the lattice constant, an estimate that can be reconciled with the data. One may speculate further that the changes in electronic DOS affect the electron-phonon coupling constant and result in faster carrier relaxation.

In summary, we have excited monolayer and bilayer graphene at MIR wavelengths between 4\,$\mu$m and 9\,$\mu$m, both on- and off-resonance with the in-plane E$_{\text{1u}}$ mode in bilayer graphene, and probed the response of the electronic structure with tr-ARPES. We find that both the peak electronic temperature as well as the relaxation rate are significantly perturbed when the excitation is made resonant with the E$_{\text{1u}}$ mode, an effect that is absent in monolayer graphene, in which light cannot couple to the in-plane lattice vibration. We explain the data by discussing the dynamical band structure changes combined with a non-adiabatic temporal response, a genuinely new type of carrier excitation for the solid state. Similar concepts may be extended beyond graphene, for example to the transition metal dichalcogenides, opening up new avenues for electronic structure control with light. The complex circular motion of the electrons throughout momentum space, which involves ultrafast manipulation of the electronic structure of the system over very few femtoseconds, may be interesting for applications in optoelectronics at very high bit rates.

\section{Acknowledgments}

We thank J{\"o}rg Harms for assisting with the figures and Axel K{\"o}hler for hydrogen-etching and argon-annealing of the samples. Access to the Artemis facility at the Rutherford Appleton Laboratory was funded by STFC. A.L. acknowledge financial support from the German Science Foundation (DFG, SFB 925) and the EU-Flagship Graphene.

\pagebreak

\section{Supplemental Material}

\subsection{Sample preparation}

Prior to graphene growth, the SiC substrate was hydrogen-etched to remove scratches from mechanical polishing, resulting in atomically flat terraces. The substrate was graphitized in argon atmosphere, and the resulting carbon layer(s) was (were) decoupled from the substrate by hydrogen intercalation. After characterization with angle-resolved photoemission spectroscopy, the samples were transported under ambient conditions to the ARTEMIS facility, where they were reinserted into ultra-high vacuum and cleaned by annealing at 200$^{\circ}$C.

\subsection{Tr-ARPES experiments}

The setup consists of a Ti:Sapphire laser system (780\,nm, 1\,kHz, 30\,fs, 15\,mJ), where part of the light is converted to MIR frequencies using an optical parametric amplifier (OPA) with difference frequency generation (DFG), resulting in wavelength-tunable pump pulses with a spectral width of $\sim$15\%. About 1\,mJ of energy is focused on an argon gas jet for high-order harmonics generation and passed through a time-preserving grating monochromator to select a single harmonic at $\hbar\omega_{\text{probe}}=31$\,eV. The extreme ultra-violet (XUV) probe ejects photoelectrons from the sample whose energies and momenta are determined by a hemispherical analyzer (SPECS, Phoibos 150), giving direct access to the occupied part of the electronic band structure.

\subsection{Frozen phonon calculations}

First-principles density functional theory calculations were performed using the Vienna \textit{ab initio} simulations package (VASP) \cite{kresse_vasp} with projector augmented (PAW) plane waves \cite{PAW1,PAW2}. The local density approximation (LDA) was employed to the exchange-correlation potential, and the Brillouin zone was sampled by a 45$\times$45$\times$1 k-mesh using the tetrahedron method with Bl\"ochl corrections \cite{blochl1994improved}. A kinetic cut-off energy of 928\,eV was employed. The LDA-optimized lattice parameters include a lattice constant of $a=2.445$ \AA\ and an interlayer spacing of 3.29 \AA. 

\subsection{Estimate displacement of carbon atoms}

The electric field amplitude $E_0$ is related to the fluence $F$ via 
$$E_0=\sqrt{\frac{2F}{c\epsilon_0\Delta t}},$$
where $c$ is the speed of light, $\epsilon_0$ is the vacuum permittivity, and $\Delta t$ is the pulse duration. The fluence in the present experiment was 0.26 mJ/cm$^2$. This corresponds to a field amplitude of $E_0$ = 1\, MV/cm for a pulse duration of $\Delta t$ = 200\,fs. The real part of the optical conductivity at resonance with the E$_{\text{1u}}$ phonon mode at $\omega_0$ = 1594\,cm$^{-1}$ for a hole concentration of $n=4\times 10^{12}$\,cm$^2$ as determined by static ARPES is $\sigma_1(\omega_0)\sim250$\,$\Omega^{-1}$cm$^{-1}$ \cite{Kuzmenko_2009}. From there, we calculate the polarization $$P=\frac{\sigma_1(\omega_0)}{\omega_0}E_0=6\times10^{-6}\text{\,C\,cm}^{-2}.$$

The polarization arises due to a light-induced dipole moment $P=d\times n\times Z_{\text{eff}}$, where $n$ is the number of dipoles per unit volume. The effective charge $Z_{\text{eff}}=0.7e$ is taken from \cite{Kuzmenko_2009}. For bilayer graphene, there are two dipoles per unit cell with volume $V=17.66$\,\AA$^3$. This results in $d=P/(n\times Z_{\text{eff}})=5$\,pm which corresponds to 2\% of the in-plane lattice constant.

\subsection{Tr-ARPES data for different pump wavelengths in monolayer and bilayer graphene}

\begin{figure}
	\center
  \includegraphics[width = 1\columnwidth]{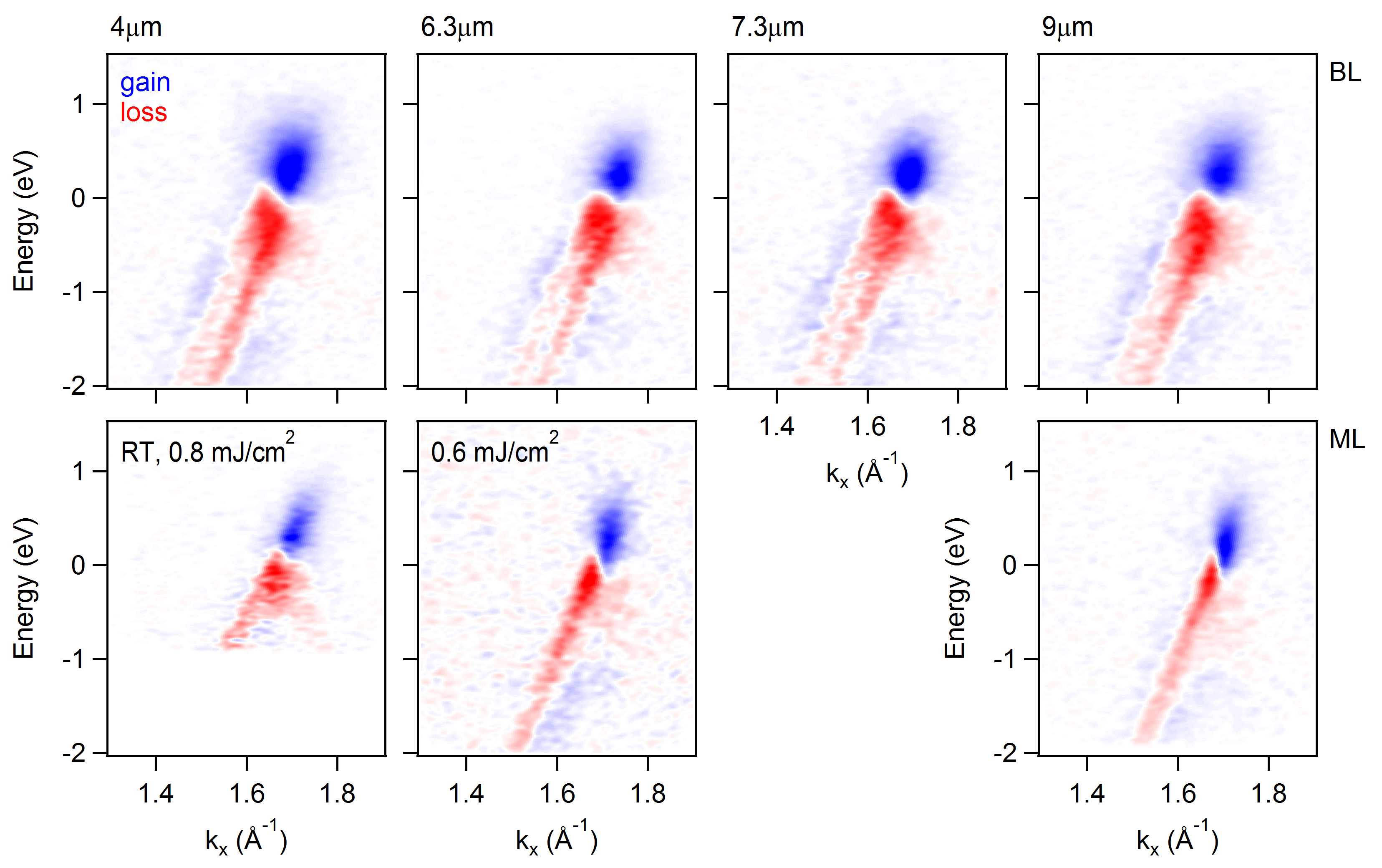}
  \caption{Peak pump-probe signal for bilayer (top row) and monolayer graphene (bottom row) for different pump wavelengths. Unless indicated otherwise, the sample temperature was 30\,K and the pump fluence 0.26\,mJ/cm$^2$.}
  \label{figS1}
\end{figure}

\begin{figure}
	\center
  \includegraphics[width = 0.9\columnwidth]{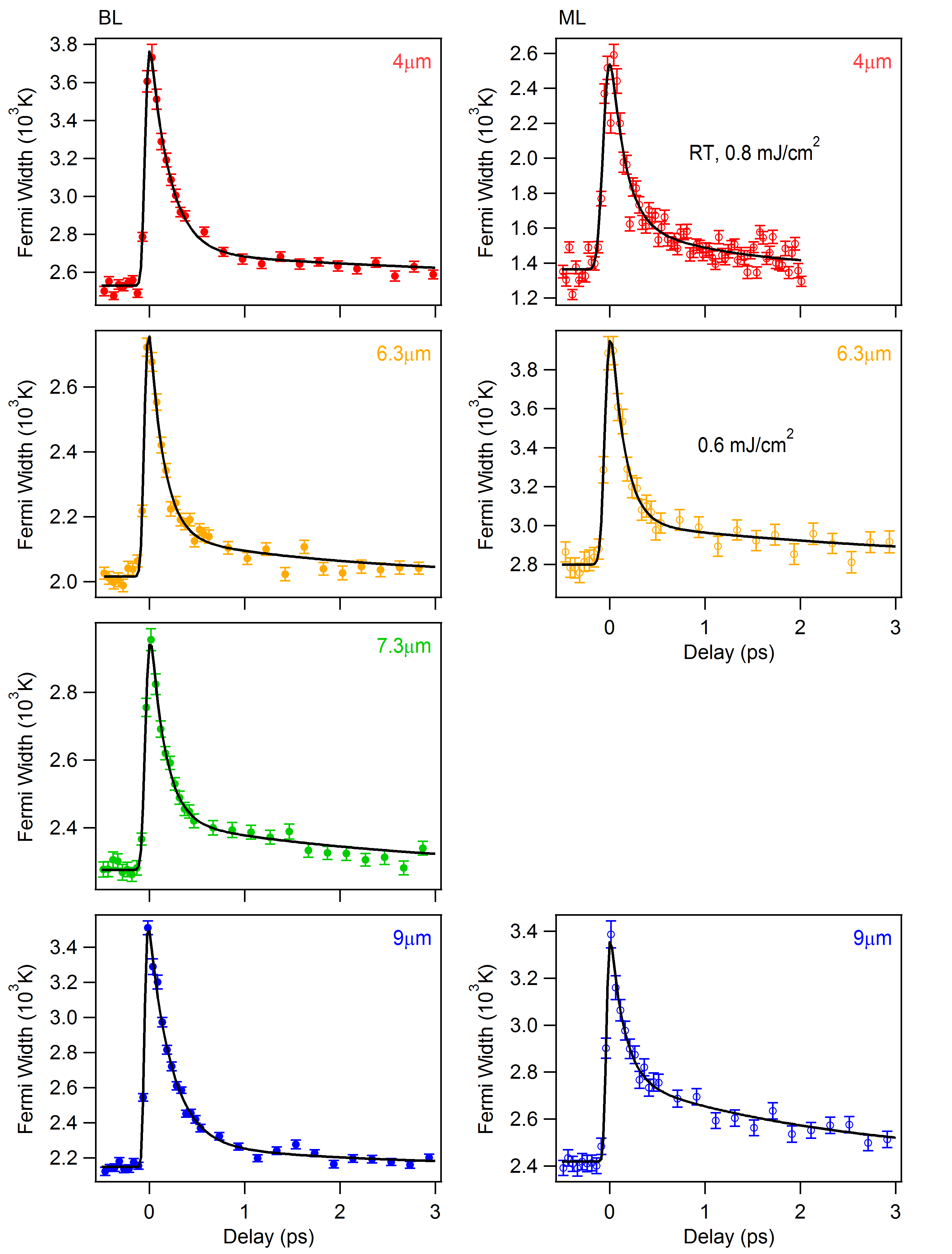}
  \caption{Width of the Fermi cut-off for bilayer (left) and monolayer graphene (right) for different pump wavelengths. Continuous lines represent fits including an error function for the rising edge and the sum of two exponentials for the decay. Unless indicated otherwise, the sample temperature was 30\,K and the pump fluence was 0.26\,mJ/cm$^2$.}
  \label{figS2}
\end{figure}

\begin{figure}
	\center
  \includegraphics[width = 0.7\columnwidth]{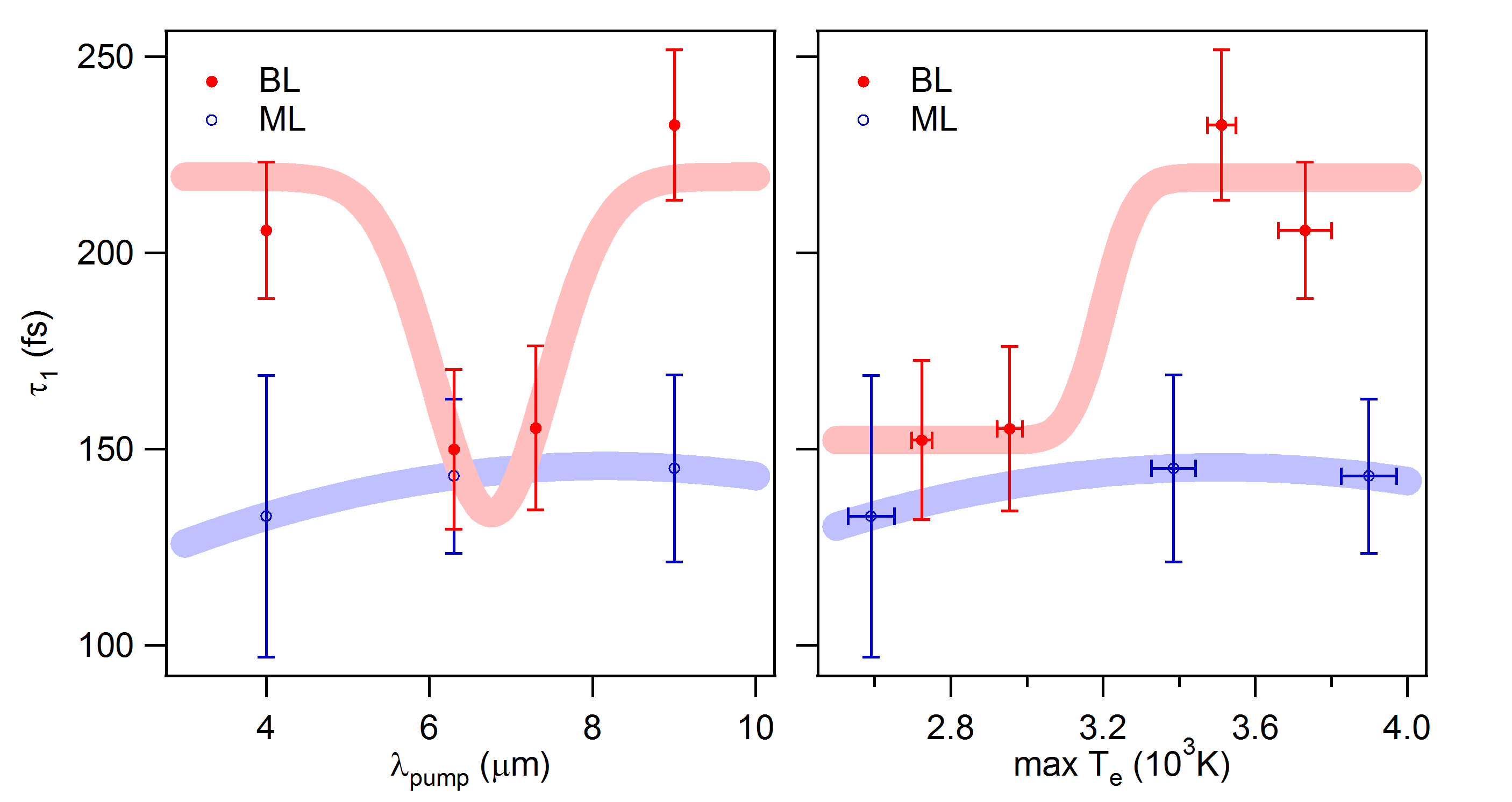}
  \caption{Comparison of the fast relaxation time $\tau_1$ in monolayer (blue) and bilayer graphene (red) as a function of pump wavelength (left) and peak electronic temperature (right). Continuous red and blue lines are guides to the eye.}
  \label{figS3}
\end{figure}

Figure \ref{figS1} shows the pump-induced changes of the photocurrent at t = 0\,fs for both bilayer (upper panel) and monolayer graphene (lower panel) for different pump wavelengths. The pump-probe signal is determined by a change in the electronic temperature and a transient broadening of the electronic structure. Figure \ref{figS2} shows the corresponding evolution of the Fermi width together with fits including an error function for the rising edge and the sum of two exponentials for the decay. The relevant fit parameters are shown in Fig. 3 of the main manuscript. In Fig. \ref{figS3} we compare the pump wavelength and electronic temperature dependence of the fast relaxation time $\tau_1$ in monolayer and bilayer graphene. The absence of any wavelength or temperature dependence in the monolayer, where the light cannot couple to the in-plane lattice vibration, indicates that the observed wavelength dependence in the bilayer has to be attributed to coherent phonon oscillations. In particular, Fig. \ref{figS3} clearly shows that we can exclude a heating effect \cite{Kemper_2014} as possible origin of the observed pump wavelength dependence.

\subsection{Electronic temperature minimum}

\begin{figure}
	\center
  \includegraphics[width = 0.8\columnwidth]{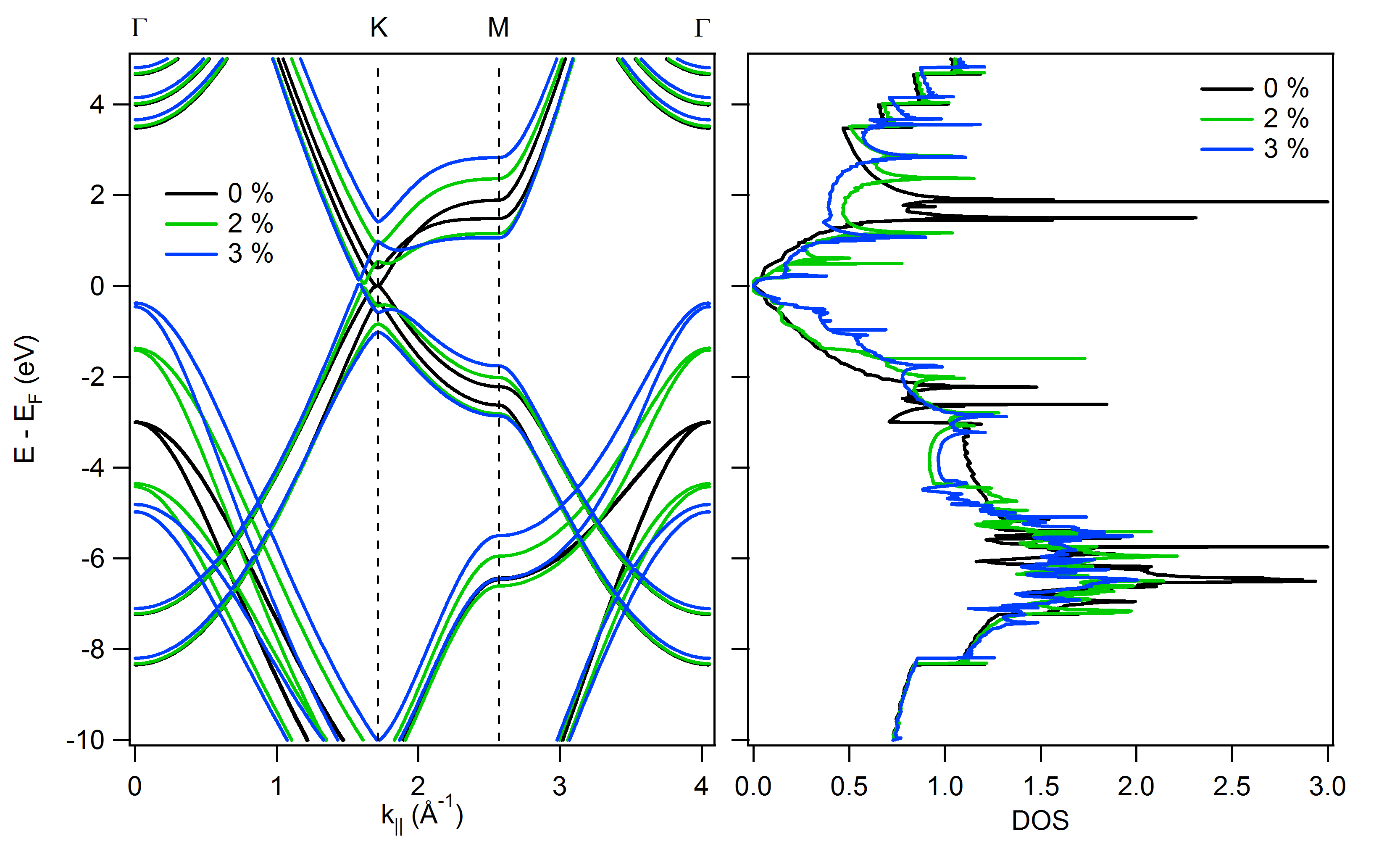}
  \caption{Calculated electronic structure (left) and density of states (right) for different lattice distortions along the E$_{\text{1u}}$ mode coordinate.}
  \label{figS4}
\end{figure}

Here, we calculate the change in electronic temperature due to a lattice distortion along the E$_{\text{1u}}$ phonon coordinate. One can see from Fig. \ref{figS3} that a typical electron-lattice relaxation process takes $\tau>0.1$\,ps. The phonon frequency is $\nu=48$\,THz. Thus $\nu\tau>1$, indicating that, at early times, the electronic subsystem can be considered to be isolated. Therefore, both the electron number $N$ and the entropy of the system $S$ are conserved during the lattice distortion. This results in the following system of coupled equations
\begin{align*}
N_0(\text{DOS}_0,\mu_0,T_0)&=N_1(\text{DOS}_1,\mu_1,T_1),\\
S_0(\text{DOS}_0,\mu_0,T_0)&=S_1(\text{DOS}_1,\mu_1,T_1),
\end{align*}
where the particle number and the entropy can be obtained from
\begin{align*}
N&=\int \text{DOS}(E) f(E) dE,\\
S&=\int \text{DOS}(E)\left[f(E)\ln f(E)+\left(1-f(E)\right)\ln\left(1-f(E)\right)\right]dE.
\end{align*}
Subscript 1 (0) refers to the (un)distorted lattice, $f(E)$ is the Fermi-Dirac distribution. The above equation is only valid if the electronic system follows a thermal distribution. This is not necessarily the case for the non-adiabatic excitation mechanism discussed in the main text. The experimental time resolution, however, is too slow to resolve this non-adiabatic dynamics, so that the measured photocurrent can be nicely fitted by a Fermi-Dirac distribution at all times. Thus, in order to compare the frozen phonon calculations to the experimental data, we assume that the electronic system can be assigned a temperature $T$. The calculated band structure and density of states for different lattice distortions are shown in Fig. \ref{figS4}. With $\mu_0$, T$_0$, and DOS$_{0,1}$ known from tr-ARPES data and frozen phonon calculations, respectively, we can easily calculate $\mu_1$ and $T_1$ for a given lattice distortion. Using $\mu_0=-200$\,meV and T$_0$ = 3600\,K, we find $\mu_1=-140$\,meV and T$_1$ = 2745\,K for an atomic displacement of 3\% of the lattice constant, in good agreement with our data.

\end{document}